\documentclass[12,leqno]{article}

\usepackage{graphics}

\def\m#1{$#1$}

\def\sgn{\;{\rm sgn}\;}
\newcommand{\beq}{\begin{equation}}
\newcommand{\eeq}{\end{equation}}

\newcommand{\beqs}{\begin{eqnarray}}
\newcommand{\eeqs}{\end{eqnarray}}

\newcommand{\DOE}{This work was supported in part by U.S.Department
of Energy grant No. DE--FG02-91ER40685}

\newcommand{\half}{\frac{1}{2}}
\newcommand{\eps}{\epsilon}

\begin{document}
\input{epsf}

\centerline{\bf\large An Interacting Parton Model for Quark and Anti-quark
distributions in the Baryon}

\centerline{V. John\footnote{vjohn@pas.rochester.edu},
G. S. Krishnaswami\footnote{govind@pas.rochester.edu} and
S. G. Rajeev\footnote{rajeev@pas.rochester.edu}}

\begin{center}
   {\it Department of Physics and Astronomy, University of Rochester,
    Rochester, New York 14627} \\
   \vspace{.5cm}

Date: June 10, 2000. Published in Physics Letters B 487 (2000)
125-132.

   \vspace{2cm}
   {\large\bf Abstract}
\end{center}

In this paper we study a 1+1 dimensional relativistic parton model for the
structure of baryons. The quarks and anti-quarks interact through a linear
potential. We obtain an analytic formula for the isospin averaged valence
quark distribution in the chiral and large $N_c$ limits. The
leading \m{1 \over N_c} and non-zero current quark mass
corrections are estimated. Then we extend this model to
include `sea' and anti-quarks. We find that the anti-quark content is
small at a low value of \m{Q^2}. Using these distributions as
initial conditions for $Q^2$ evolution, we compare with experimental
measurements of the structure function \m{xF_3(x,Q^2)} and find reasonable
agreement. The only parameters we can adjust are the fraction of baryon
momentum carried by valence quarks and the initial scale \m{Q_0^2}.



\pagebreak

\section{Introduction}

    We present a variational parton model description for the
structure of baryons as measured in Deep Inelastic
Scattering. This model enables us to calculate the \m{x_B} dependence of
the structure function \m{xF_3} at an initial value of
\m{Q_0^2}. We compare this prediction with experimental measurements by
CCFR and CDHS collaborations.

In Deep Inelastic Scattering, the longitudinal momenta of the partons
dominates their transverse momenta. Indeed, as pointed out by Altarelli,
Parisi and others \cite{dglap}, a perturbative treatment of transverse
momenta, with an upper cut-off $Q$, leads to the same scaling violations
as predicted by the Operator Product Expansion in the leading logarithmic
approximation. By the uncertainty principle, the virtual photon momentum
$Q$, is a measure of the size of transverse momenta being probed.

While the $Q^2$ dependence (for sufficiently large $Q^2$) of the
structure functions is well understood \cite{dglap,cteqmrst},
the $x_B$ dependence is harder to understand since it deals with the
formation of a relativistic bound state. We make the following
ansatz: At some low value of $Q^2 = Q_0^2$,
the transverse momenta of the partons may be ignored as a first approximation.
The $x_B$ dependence of quark
and anti-quark distributions at $Q_0^2$, is then determined by solving a 1+1
dimensional model. In this model, quarks interact via a linear
potential in the null coordinates. This is the simplest potential consistent
with Lorentz covariance.

We also suppose that the valence quarks carry a fraction $f$ of the total
baryon momentum at $Q^2 = Q_0^2$. The rest being carried by gluons,
anti-quarks and sea-quarks. In a previous numerical study \cite{ipm},
we determined the valence quark distribution in this interacting parton model.
Here, we extend this model to include anti-quarks and also obtain analytic
formulae for the valence quark distributions, within a variational approach.
Based on the two parameters $Q_0$ and $f$, we predict the initial $x_B$
dependence of the iso-spin averaged quark and anti-quark distributions.
Finally, to compare with experimental data at higher $Q^2$, we use the
solutions of this model as initial conditions for the DGLAP evolution
equations.

We do not derive this model from a more basic theory. It is proposed merely
as a phenomenological parton model for an approximate description of
Deep Inelastic Scattering.

Let us now give a brief introduction to our analysis of the interacting
parton model. The partons are assumed to be relativistic particles
interacting through a linear potential. The number of colors
\m{N_c} is kept variable and we work mostly in
the limit of a large number if colors. To simplify this many body
problem we ignored the anti-quark degrees of freedom in \cite{ipm}.
The baryon wave function was determined by the principle that
it minimizes the \m{(mass)^2} of the baryon.
Within a Hartree approximation, the valence quark wavefunction
is the solution of a non-linear integral equation which was
solved numerically in \cite{ipm}.

In this letter we first show that the true minimum of \m{(mass)^2} occurs
for a configuration that consists only of valence quarks,
in the chiral and large \m{N_c} limits.
The deviation of the anti-quark distribution from zero is measured by the
dimensionless parameter \m{m^2 \over {\tilde g}^2} in addition to
\m{1 \over N_c} corrections, which we estimate. Here \m{m} is the current
quark mass and \m{\tilde g} a coupling constant.

In order to determine the anti-quark content of the baryon, we perform a
unitary transformation on the Fermionic Fock space, starting from
a purely valence state. This transformation is like a Bogoliubov
transformation which mixes positive and negative momentum states.
It is sufficient to consider Bogoliubov transformations parametrized by
a single angle \m{\theta} which is determined by minimizing the
\m{(mass)^2} of the baryon. We find that in the
large $N_c$ limit, \m{\theta} vanishes for
zero current quark mass. For physically reasonable values of
${m^2 \over \tilde{g^2}}$, the anti-quarks carry
less than a percent of baryon momentum. This is at an initial value
of \m{Q^2 = Q_0^2}, at which transverse momenta are neglected.
The isospin averaged quark and anti-quark distributions
are found within a variational approximation.

To compare our results with experimental data,
we evolve the distributions to higher values of \m{Q^2}
via the DGLAP equations \cite{dglap}. However, the gluon distributions,
which we have not determined, appear in the evolution equations. It turns
out that the difference between quark and anti-quark distributions
(valence quark distribution: \m{q^V(x_B,Q^2) =
\sum_{\beta = u,d}(q^{\beta}(x_B,Q^2)-q^{\bar \beta}(x_B,Q^2))})
evolves independently of the gluon
distribution to leading order. Moreover, if we ignore certain correlations,
this difference is the average of the structure function \m{F_3}
measured in neutrino and anti-neutrino Deep Inelastic Scattering
\cite{cteqmrst}. Given the \m{x_B} dependence of the parton
distribution functions (PDFs) at an initial \m{Q_0^2}, their
\m{Q^2} evolution is determined by the DGLAP equation with the
splitting function \m{P_{qq}} calculated perturbatively:

\begin{center}
\m{ {dq^V(x_B,t) \over dt} = {\alpha_s(t) \over 2\pi}
\int_{x_B}^{1} {dy \over y} q^V(y,t) P_{qq}({x_B \over y}).}
\end{center}

\noindent Here \m{t = \log(Q^2/Q_0^2)}.
The normalization \m{\nu(Q_0^2) = \int_0^1 dx_B q^V(x_B,Q_0^2)} is
determined by integrating the DGLAP equation with
initial condition \m{\nu(\infty) = N_c = 3} from \m{Q^2 = \infty} to
\m{Q_0^2}. Due to the large \m{Q^2} range involved,
 we determine \m{\nu(Q_0^2)} to high order.
Within our approximations, \m{\nu(Q_0^2)~=~\int_0^1 dx_B F_3(x_B,Q_0^2)},
which is given by the GLS sum rule \cite{gls}. If we denote the
isospin averaged valence quark probability density as \m{V(x_B,Q_0^2)}, then
\m{q^V(x_B, Q_0^2)~=~\nu(Q_0^2)~V(x_B,Q_0^2)}. In section $4$ we compare
our predictions for \m{xF_3(x,Q^2)} with experimental measurements
by the CDHS and CCFR collaborations. Our predictions
agree well with data for a choice of parameters \m{f = \half}
and \m{Q_0^2 = 0.4} GeV$^2$ for which \m{\nu(Q_0^2) = 2.25}. This
choice of parameters is consistent
with phenomenological fits to data \cite{cteqmrst,grv}. However, it would
be useful to know what the `best-fit' values of these parameters are.

An impressive discretized light-cone (DLCQ) analysis of 2 dimensional QCD
was done by Hornbostel et. al. \cite{hornbostel}. Our phenomenological
parton model provides a complementary physical approach to their
more direct numerical diagonalization of the hamiltonian. The
2-dimensional valence quark wave function we find reduces precisely
to the one obtained in \cite{hornbostel} when we set \m{f = 1}.
We do not find a similar concordance  with the more conventional
lattice QCD methods: the lightcone methods seem to incorporate the
physical phenomena much more directly. Other DLCQ \cite{dalley}
calculations study the meson and glueball
spectra of 2d models. We focus on the baryon. For other approaches see for
instance \cite{brodskyetal}.


\section{Valence Parton Model}

Let us begin by reviewing the valence quark approximation.
Ref. \cite{ipm} may be consulted for details.
We assume that the momenta of the partons
in the \m{x^1} direction are large compared to the transverse
momenta; which we ignore. We use null co-ordinates \footnote{See appendix
to \cite{2dqhd} for kinematics.} where the null momentum
\m{p = p_0 - p_1} is the basic kinematic variable.
Then the kinetic energy of a free particle of mass $m$ is
\m{p_0 = \half (p + {m^2 \over p})}. So the wave function of a quark
will vanish for negative \m{p} while that of an anti-quark vanishes for
positive $p$.

    If we ignore anti-quarks as in \cite{ipm}, then the
baryon wavefunction \m{\tilde \psi(\{\nu_i,\alpha_i,p_i\})}
depends on the colors, flavours \m{(\alpha_i = 1,\cdots,M)}
and null momenta of the \m{N_c}
valence quarks. \m{\tilde \psi(\{\nu_i,\alpha_i,p_i\}) =
\eps_{\nu_1,\cdots,\nu_{N_c}} \tilde \psi(\{\alpha_i,p_i\})}
since the baryon is a color singlet.
Moreover, since the null momenta are positive,
the sum of quark momenta cannot exceed the total baryon momentum \m{P}.
In particular, the wave function must vanish for \m{p_i > P}.
Since the \m{\eps} tensor is anti-symmetric in color, the wavefunction
must be symmetric in the remaining variables: partons behave like
bosons in the momentum, spin and flavour variables. The ground state
wave function is determined by minimizing the total energy

\begin{center}
\m{ {\cal E}_{N_c}[\tilde\psi] = \sum_{\alpha_1\cdots
\alpha_{N_c}}\int_0^P \sum_{i=1}^{N_c}\half[p_i+{\mu_{\alpha_i}^2\over
p_i}]|\tilde\psi(\alpha_1,p_1;\cdots \alpha_{N_c},p_{N_c})|^2{dp_1\cdots
dp_{N_c}\over (2\pi)^{N_c}}}

\hspace{.2in} \m{ + \half g^2 \sum_{\alpha_1\cdots \alpha_{N_c}}
\int_{-\infty}^\infty \sum_{i\neq j} v(x_i-x_j)|\psi(\alpha_1,x_1;
\cdots \alpha_{N_c},x_{N_c})|^2 dx_1\cdots dx_{N_c}.}
\end{center}

\noindent Here \m{\mu_{\alpha_i}^2=m_{\alpha_i}^2 - {{\tilde g}^2\over \pi}}
is the effective mass of the parton \cite{ipm}, which avoids a potential
infrared divergence in the potential energy. Also, \m{{\tilde g}^2 =
{g^2 \over N_c}}. The simplest potential consistent with Lorentz invariance
is linear, \m{v(x) = {|x| \over 2}}, which is also favoured by phenomenology
\cite{potmodel}. \m{\tilde g} is a coupling constant with the dimensions of
mass. Our predictions turn out to be independent of $\tilde g$ in the chiral
limit. In the ground state, we expect the Hartree ansatz

\begin{center}
\m{ \tilde\psi(\{\alpha_j,p_j\})=
2\pi\delta(P-\sum_ip_i)\prod_{i=1}^{N_c}\tilde\psi(\alpha_i,p_i).}
\end{center}

\noindent to be a good approximation. The valence quark wave function
is normalized to have unit length,
\m{\sum_{\alpha=1}^M\int_0^P|\tilde\psi(\alpha,p)|^2 {dp\over 2\pi}=1}.
It must also satisfy the momentum sum rule:
\m{N_c~\sum_{\alpha}~\int_0^P~p|\tilde\psi(p)|^2{dp\over 2\pi}~=~fP.}
\m{f} here is the fraction of baryon momentum carried by the valence
quarks, which is roughly a half at low \m{Q^2} \cite{cteqmrst}. In these
two formulae, we are ignoring correlations that are suppressed
for large-$N_c$. They differ from the exact formulae
in the same way as the canonical ensemble differs from the
micro-canonical ensemble in statistical mechanics.

Since we are interested in the isospin averaged distributions,
we will average over spin-flavour degrees of freedom. Thus we look for
a wave function that is non-zero only for
a single value of \m{\alpha}: \m{\tilde\psi(\alpha,p) =
\delta_{\alpha,1}\tilde\psi(p)}.

In second quantized language, this corresponds to the valence state
\m{|V> = a^{1 \dag}_{\tilde{\psi}} \cdots a^{N_c \dag}_{\tilde{\psi}} |0>}.
Here, \m{a^{j \dag}_{\tilde \psi}} creates a quark
with color \m{j} in the state \m{\tilde \psi}. These operators
satisfy canonical anti-commutation relations (CAR):
\m{ \{ a_{i u},a^{j \dag}_{v} \}
= \delta_{i}^{j} <u,v> } with respect to the Dirac vacuum \m{|0>}
where all negative energy states are filled and positive ones empty:
\m{a_{\tilde {\psi}_-}^{i^{\dag}} |0> = 0} and
\m{a_{j \tilde {\psi}_+} |0> = 0}.
\m{\tilde {\psi}_-(p)} vanishes for \m{p \geq 0} and
\m{\tilde {\psi}_+(p)} for \m{p \leq 0}. The Pauli principle
requres that the density matrix \m{\tilde \rho_V(p,q) =
<~V~|~{1 \over N_c}\hat a^{i \dag}(p) \hat a_i(q)|V>}
is a hermitean projection operator: \m{\int_{-\infty}^{\infty}
\tilde \rho_V(p,r)~\tilde \rho_V(r,q)~{dr \over 2\pi}~=~\tilde \rho_V(p,q)}.
The eigenvalues of the density matrix are the occupation numbers
of particles, for a projection operator these are $0$ or $1$ as required
by the Pauli principle. For a state containing one baryon,
the normal ordered trace of the density matrix is equal to one :
\m{tr(\tilde \rho_V(p,q)~+~\half \tilde\delta(p,q)(sgn(p)-1))~=~1}
by a use of the CAR. \m{\tilde \delta(p,q)}
is the identity matrix. The above Hartree ansatz,
\m{\tilde \rho_V(p,q)~=~\tilde\psi(p) \tilde\psi^{*}(q) +
\half \tilde \delta(p,q)(1-\sgn{p})}
satisfies these constraints.

A Lorentz invariant formulation is to minimize the
mass \m{{{\cal M} \over N_c}} of the Baryon per quark:

\vspace{.1in}
\m{
{{\cal M}^2 \over \tilde g^2 N_c^2}=\left[\half \int_0^P p|\tilde
\psi(p)|^2 {dp\over 2\pi}\right]* }

\vspace{.1in}

\hspace{.1in}\m{ \left[{1\over 2} \int_0^P {1 \over 2p}
|\tilde\psi(p)|^2 ({m^2 \over \tilde g^2} - {1 \over \pi}) {dp\over 2\pi}
+ \half \int_{-\infty}^{\infty} dxdy |\psi(x)|^2  |\psi(y)|^2
\half |x-y| \right]}
\vspace{.1in}


\subsection{Analytic results in the large \m{N_c} limit}

    In \cite{ipm} we solved the integral equation for the minimization of
energy numerically. There is in fact an analytic solution in the chiral
\m{({m^2 \over \tilde g^2} \to 0)} and large $N_c$ limits.
The boundary condition is that $\tilde\psi(p)$ must vanish for $p>P$. However,
$P$ is an extensive variable, $P \sim N_c$. So for $N_c = \infty$, the valence
quark wave function is not required to vanish for any finite value of $p$.
If we use the intensive quantity $\bar P = {P \over N_c}$, the analog of
momentum fraction is $\bar x_B = {p \over \bar P}$, but the wave function is
not required to vanish beyond $\bar x_B = 1$. In order to compare directly
with a wave function computed for $N_c = 3$, we pick the reference frame
in which $\bar P = {1 \over 3}$. The momentum sum rule then becomes
$\int_0^{\infty} p |\tilde\psi(p)|^2 {dp \over 2\pi} = f \bar P$.
It can be checked explicitly that
\m{\tilde \psi(p) = \sqrt{2\pi \over f \bar P} e^{-p \over 2f \bar P}}
is an exact solution to the integral equation for the minimization of
baryon \m{(mass)^2}. Alternatively, we can calculate
\m{{\cal M}^2 \over N_c^2} for
this wavefunction and see that it is zero. The potential
and self energies cancel each other. Thus, in this limit
 the minimum of \m{(mass)^2} actually occurs for
a purely valence quark configuration. Since \m{\tilde g \sim \Lambda_{QCD}
\sim 200} MeV and $m_u$, \m{m_d \sim 5-8} MeV, this should be a good
approximation provided the \m{1 \over N_c} corrections are small.
This is indeed the case, as we show below. For $f = \half$, the valence
quark density normalized to one is \m{V(\bar x_B) = 6e^{-6 \bar x_B}}.


\subsection{Leading order \m{1 \over N_c} correction}

The leading order effect of finite \m{N_c} is to restrict the range of
quark momenta to \m{p < P}. Now, \m{(1- {p \over n})^n \to e^{-p}}
as \m{n \to \infty}. Therefore, \m{\tilde \psi(p) = C p^a (1-{p \over P})^b}
, \m{0 \leq p \leq P} should be a good ansatz for the ground state wave
function for finite \m{N_c}. $C$ is determined by normalization.
The momentum sum rule implies that \m{b = {N_c \over 2f} - 1
+ a({N_c \over f} - 1)}. The minimization of energy implies that
\m{a} satisfies the transcendental equation

\begin{center}
\m{ {\pi m^2\over \tilde g^2}=1+\int_0^1{dy\over y^2}\bigg[
 (1+y)^{a}+ (1-y)^{a}-2\bigg]+
\int_1^\infty {dy\over y^2}\bigg[(1+y)^a-2\bigg].}
\end{center}

\noindent which we derived in \cite{ipm}. In the limit of chiral
symmetry, \m{a \to 0}. If valence quarks carry all the momentum of the
baryon, \m{f = 1}, and \m{V(x_B) = (N_c-1)(1-x_B)^{N_c-2}}, which is identical
to the result obtained from DLCQ, reported in \cite{hornbostel}.
However, valence quarks carry only about half the baryon momentum,
so that for \m{N_c = 3} our variational estimate for the valence quark
density is \m{V(x_B,Q_0^2) = 5(1-x_B)^4}; this agrees well with our numerical
solution from Ref. \cite{ipm}.

\begin{figure}
\centerline{\epsfxsize=6.truecm\epsfbox{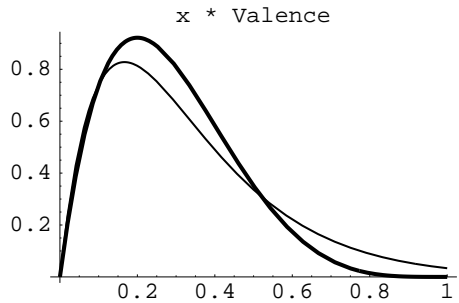}}
\caption{ Comparison of valence quark distribution
\m{\bar x_B q^V(\bar x_B,Q_0^2)} obtained in the large-\m{N_c} limit
(thin curve) with the variational estimate \m{x_B q^V(x_B,Q_0^2)}
after taking into account the leading \m{1 \over N_c} correction
(thick curve).}
\end{figure}

It is evident from Fig. 1 that the primary effect of the
\m{1 \over N_c} correction is to make the distribution vanish beyond
${p \over P} = 1$. The actual shape of the distribution is
already well captured by our analytic solution in the large-$N_c$ limit.


\section{Extension of Parton model to include Anti-quarks}

Now we turn to the anti-quark content of the baryon.
As we will see below, the minimum of Baryon \m{(mass)^2} occurs for a
small anti-quark content when the parameter \m{m^2 \over \tilde g^2}
is small. In order to determine it, we need to allow for states with
negative momenta. However, only the difference between the quark and
anti-quark numbers is conserved in the full theory.
Therefore, the baryon must be in a linear superposition of states
containing $\eta$ anti-quarks and $N_c+\eta$ quarks,
for \m{\eta = 0,1,\cdots,\infty}. The energy of a state containing
\m{\eta} anti-quarks is

\vspace{.1in}
\hspace{.1in} \m{ \sum_{\nu_j} \sum_{i=1}^{N+2\eta} \int_{-P}^{P}
\half (p_i + {\mu^2 \over p_i}) |\tilde \psi_\eta (\{\nu_j,p_j\})|^2
\Pi_k dp_k}

\vspace{.1in}

\hspace{.2in} \m{+ {g^2 \over 2} \sum_{\nu_k} \sum_{i \ne j}
\int_{-\infty}^{\infty}
\half |x_i - x_j| |\psi_\eta(\{\nu_k,x_k\})|^2 \Pi_m dx_m }
\vspace{.1in}

\noindent where \m{\tilde \psi_\eta(\{\nu_i,p_i\})} is the wave
function of such a state.
\m{\{\nu_i,p_i\}_{i=1}^{N_c+\eta}} refer to the colors and
momenta of the quarks, while \m{\{\nu_i,p_i\}_{i=N_c+\eta+1}^{N_c+2\eta}}
refer to anti-quarks. The wave function vanishes for negative quark
momenta and positive anti-quark momenta. The total energy is the sum
of the energies for each value of $\eta$.

Now we shall argue that the {\it ground state}
of the baryon is to a good approximation determined by three
orthonormal one-parton states \m{\psi} and
\m{\psi_+} which describe quarks and \m{\psi_-}
which describes anti-quarks. We will continue to work in a
factorized Hartree approximation, ignoring correlations
except when they are required by the Pauli principle or color invariance.

In the absence of anti-quarks, there are just $N_c$ quarks whose
wavefunction is completely anti-symmetric in color. Within
the Hartree approximation, they all occupy the same single parton
positive momentum state \m{\psi}, which minimizes
the \m{(mass)^2} of the baryon. Now consider adding a quark and an
anti-quark ($\eta = 1$). The anti-quark will occupy the negative
momentum state \m{\psi_-} that minimizes the energy. Suppose
all the \m{N_c + 1} quarks occupy the state \m{\psi}. Then, by the
Pauli exclusion principle, the color part of the quark wave
function must be totally
anti-symmetric. However, there is no completely anti-symmetric tensor
in $N_c + 1$ indices transforming under $SU(N_c)$. Therefore, we are forced to
introduce a new positive momentum state \m{\psi_+},
that must be orthogonal to the filled state \m{\psi}. In order that
this cost a minimal amount of energy, we expect \m{\tilde \psi_+(p)} to
have only one more node than \m{\tilde \psi(p)}. As long as \m{\eta \leq N_c},
in the ground state, the quarks and anti-quarks occupy the states
\m{\psi, \psi_+, \psi_-}. If \m{\eta > N_c}, we
would have to introduce another pair of orthonormal states. However,
we find that these additional corrections are very small.

Now we shall argue that the configuration containing valence, sea and
anti-quarks (say \m{|VSA>})  can be obtained by a unitary
transformation acting on the valence quark state \m{|V>}: a Bogoliubov
 transformation. It must be unitary in order that \m{<VSA|VSA>=1}.
The operator that creates a quark in state
\m{\psi_+} and an antiquark in state \m{\psi_-} is
\m{a_{j\psi_-}a^{j\dag}_{\psi_+}}; we sum over color indices to produce
a color invariant state. Thus the unitary
transformation we seek is the identity except in the two dimensional
subspace spanned by \m{\psi_+} and \m{\psi_-}. Thus our variational
ansatz is  \m{|VSA>=e^{\theta[a_{j\psi_-}a^{j\dag}_{\psi_+}-{\rm
h.c.}]}|V>}.  The density matrix of quarks in the new state can now be
calculated:

\vspace{.05in} \hspace{.2in} \m{
\tilde{\rho}_{VSA}(p,q) = \tilde\psi(p) \tilde\psi(q) -
 \sin^2{\theta} [ \tilde\psi_-(p)\tilde\psi_-(q) -
\tilde\psi_+(p) \tilde\psi_+(q) ]}

\vspace{.08in}

\hspace{.5in} \m{ - \half\sin{2 \theta} [\tilde\psi_-(p) \tilde\psi_+(q) +
\tilde\psi_+(p) \tilde\psi_-(q)]  + \half \tilde\delta(p,q) (1- \sgn{p}).
} \vspace{.05in}

\noindent Physical quantities are expressed most simply in terms
of the normal ordered density matrix:
\m{\tilde M(p,q) = - 2 \tilde{\rho}(p,q) +
\tilde\delta(p,q)(1- \sgn{p})}. For example, the baryon number
density in momentum space is

\vspace{.05in}
\centerline{\m{\tilde M(p,p)-\tilde M(-p,-p)=|\tilde\psi(p)|^2+
\sin^2{\theta}\left[|\tilde\psi_+(p)|^2-|\tilde\psi_-(-p)|^2\right].}}
\vspace{.05in}

This confirms the interpretation of  \m{\psi_-}
as the anti-quark wavefunction. The \m{(mass)^2} is given by

\vspace{.1in}
\hspace{.2in} \m{{{\cal M}^2 \over N_c^2}=\left[-\half
\int_{-P}^{P} p\tilde M(p,p){dp\over 2\pi}\right] * }

\vspace{.1in}

\hspace{.4in} \m{ \left[{-{1\over 2}} \int_{-P}^{P} \tilde M(p,p){\mu^2\over
2p}{dp\over 2\pi}+{\tilde
g^2\over 8}\int_{- \infty}^{\infty} dxdy |M(x,y)|^2\half|x-y|\right]
} \vspace{.1in}

The variational quantities \m{\psi, \psi_+, \psi_-}, and \m{\theta}
are determined by minimizing the baryon \m{(mass)^2}. In the ground state,
we expect \m{\tilde \psi(p)} and \m{\tilde \psi_-(p)} to have no nodes
(except possibly at the boundaries $p = 0, P$), while \m{\tilde \psi_+}
must have one more node. We estimate them variationally.


\subsection{Large $N_c$ analysis}

Working in the
\m{N_c \to \infty} limit, the form of the analytic solution suggests the choice
\m{\tilde\psi(p)=C\left({p\over \tilde g}\right)^a e^{-b{p\over \tilde g}},
\tilde\psi_+(p)=C_+\left({p\over \tilde g}\right)^a\left[{p\over \tilde
g}-C_1\right] e^{-b{p\over \tilde g}}} for \m{p>0} and
\m{\tilde\psi_-(p)=\tilde\psi(-p) } for \m{p<0}. (For other ranges of
\m{p} these functions must vanish.) \m{C_1} is
determined by the orthogonality condition while \m{C,C_+} are fixed by
the normalization conditions. The variational parameter \m{b}
determines the reference frame. The Lorentz invariant quantity
\m{{\cal M}^2} is independent of \m{b}. Thus the variational
principle will determine \m{a} and \m{\theta} and hence the wavefunctions.
The actual minimization of \m{{\cal M}^2} is a lengthy but straightforward
calculation. Most of the energy integrals can
be evaluated analytically and we do them using the
symbolic package Mathematica. We find that \m{\theta, a \to 0} as
\m{{m^2 \over \tilde g^2} \to 0} recovering the purely valence
exponential solution. For a physically reasonable value of
\m{{m^2 \over \tilde g^2} \sim {m_{u,d}^2 \over \Lambda_{QCD}^2} \sim 0.001},
we estimate \m{\theta = 0.02} and \m{a = 0.035}. This corresponds to a small
but non-vanishing anti-quark content in the baryon.


\subsection{Leading order \m{1 \over N_c} correction}

As in the valence quark case, the leading \m{1 \over N_c}
effect is to make these wave functions vanish beyond \m{ p = P}.
We estimate this correction using the ansatz
\m{\tilde\psi(p)=D p^a (1 - p)^b, \tilde\psi_+(p) =
D_+ p^a (p-D_1) (1-p)^b} for \m{1 \geq p \geq 0} and
\m{\tilde\psi_-(p)=\tilde\psi(-p) } for \m{-1 \leq p \leq 0}. Here
\m{P = 1} and for other ranges of \m{p} these functions must vanish.
\m{D_1} is determined by the orthogonality condition while
\m{D,D_+} are fixed by the normalization conditions.
For the choice \m{Q_0^2 = 0.4 GeV^2}, \m{f = \half},
\m{{m^2\over\tilde{g}^2} \sim~.001},
 we get  \m{\theta = 0.02}, \m{a = 0.035}, \m{b = 2.175} for the variational
 parameters. The valence quark distribution is normalized to
\m{\nu(Q_0^2) = 2.25} while the normalization of the  anti-quark
distribution is determined as a consequence to be
\m{\nu(Q_0^2)\sin^2{\theta}}. Since \m{\sin^2{\theta} \sim 10^{-4}},
the primordial anti-quarks carry only about \m{.01 \%} of the baryon
momentum.

These results are identical to what we obtained from a more field theoretic
point of view in \cite{2dqhd,xf3anti}. They also agree with the
DLCQ analysis of \cite{hornbostel} as pointed out in Section 2.2.
However, the parton model point of view presented here is much simpler.
Moreover, the GRV collaboration
\cite{grv} have obtained a reasonably good fit to Deep Inelastic Data for
\m{x_B > 10^{-2}} starting with a vanishing anti-quark distribution
at an initial \m{Q_0^2 \sim 0.2} GeV$^2$. Our approximations are not
expected to be valid for extremely low values of the momentum fraction, where
the assumption that longitudinal momenta dominate, becomes questionable.

Thus we find that the valence quark picture is quite accurate: the
`primordial' anti-quark distribution is very  small.
The  anti-quark content is zero not only in the non-relativistic
limit \m{m>>\tilde g} but also (somewhat surprsingly) in the
chiral limit \m{m=0} when \m{N_c \to \infty}, with \m{1 \over N_c}
corrections being small. Nevertheless, a substantial anti-quark content
is generated by $Q^2$ evolution.

\section{Comparison with experimental data}

Finally, we compare with experimental data at higher values of \m{Q^2}.
The DGLAP $Q^2$ evolution equation is
integrated numerically. We set \m{N_c=3}, \m{\Lambda_{QCD}=200} MeV
and the current quark mass \m{m = 0}.
The parameters \m{f, Q_0^2} should be determined by a best fit to
experimental data. For now we assign to them reasonable values
\m{f = \half} and \m{Q_0^2 = 0.4} GeV$^2$ which are consistent
with phenomenological fits to data.
From the GLS sum rule we get \m{\nu (0.4~GeV^2) = 2.25} \cite{gls}.
In Fig. 2, we show a comparison with \m{xF_3} measurements by the
CDHS and CCFR collaborations \cite{cdhsccfr} at \m{Q^2 \sim 13} GeV$^2$.
The small range \m{0.4 \leq Q^2 \leq 13} GeV$^2$ over which we are evolving
justifies the use of the leading order DGLAP equation. The plot
shows that our prediction agrees reasonably well with the experimental
measurements.

\begin{figure}
\centerline{\epsfxsize=6.truecm\epsfbox{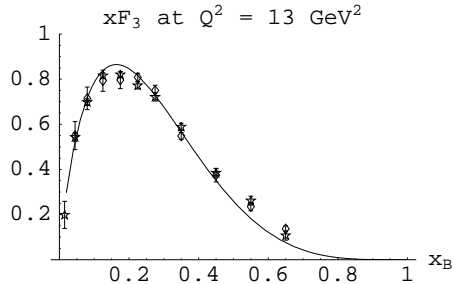}}
\caption{Comparison of predicted \m{xF_3} at
\m{Q^2=13} GeV$^2$ (solid curve) with measurements by CCFR (\m{\star}) at
\m{12.6} GeV$^2$ and CDHS (\m{\diamondsuit}) at
\m{12.05 \leq Q^2 \leq 14.3} GeV$^2$. \m{Q_0^2 = 0.4} GeV$^2$
and \m{f = \half}.}
\end{figure}

\section*{Acknowledgements}
\DOE.

\end{document}